\documentclass[aps,prd,twocolumn,superscriptaddress,nofootinbib,floatfix]{revtex4}

\usepackage[dvips]{graphicx}
\usepackage{amsmath,amssymb,latexsym}
\usepackage{bm}
\usepackage{epsfig}

\newcommand{\postscript}[2]{\setlength{\epsfxsize}{#2\hsize}
   \centerline{\epsfbox{#1}}}

\begin{document}

\title{Probing leptoquark production at IceCube}

\author{Luis A.~Anchordoqui}
\affiliation{Department of Physics,
University of Wisconsin-Milwaukee, P.O. Box 413, Milwaukee, WI 53201, USA
}

\author{Carlos A. Garc\'{\i}a Canal}
\affiliation{IFLP (CONICET) $-$ Dpto. de F\'{\i}sica,
Universidad Nacional de La Plata, C.C.67, La Plata (1900), Argentina}

\author{Haim Goldberg}
\affiliation{Department of Physics,
Northeastern University, Boston, MA 02115, USA}

\author{Daniel Gomez Dumm}
\affiliation{IFLP (CONICET) $-$ Dpto. de F\'{\i}sica,
Universidad Nacional de La Plata, C.C.67, La Plata (1900), Argentina}

\author{Francis Halzen}
\affiliation{Department of Physics,
University of Wisconsin, Madison, WI 53706, USA}

\date{September 2006}
\begin{abstract}
  \noindent We emphasize the inelasticity distribution
  of events detected at the IceCube neutrino telescope as an important
  tool for revealing new physics. This is possible because the
   unique energy resolution at this facility allows to separately assign
  the energy fractions for emergent muons and taus in neutrino
  interactions. As a particular example, we explore the possibility
  of probing second and third generation leptoquark parameter space
  (coupling and mass).  We show that production of leptoquarks with
  masses $\agt 250$~GeV and diagonal generation couplings of ${\cal
    O}(1)$ can be directly tested if the
  cosmic neutrino flux is at the Waxman-Bahcall level.
\end{abstract}


\maketitle

\section{Introduction}

Leptoquarks are $SU(3)$-colored particles which simultaneously carry
non-zero baryon and lepton quantum numbers. They are predicted in
several models (such as $SU(5)$~\cite{Georgi:1974sy} or Pati-Salam
$SU(4)$~\cite{Pati:1974yy}) addressing the unification of the lepton
and quark sectors of the standard model (SM). In such models, the
masses of the leptoquarks are generically superheavy, on the order of
the GUT scale, which puts them out of reach of direct experimental
access. Nevertheless, since leptoquarks with electroweak scale masses
are not disallowed for any fundamental reason, it is of interest to
conduct experimental searches to delimit their
properties~\cite{Dorsner:2005fq}. It is, of course, important to note
that in order to avoid rapid baryon decay, the simultaneous trilinear
coupling of the leptoquark to a purely hadronic channel needs to be
excluded~\cite{Hinchliffe:1992ad}.

In general, the couplings of the leptoquark need not be
generation-diagonal, and the problem of extracting limits on couplings and
masses is complicated by the presence of a large-dimensional parameter
space.  Experiments at HERA have placed lower limits of ${\cal O}
(300~{\rm GeV})$ on first generation leptoquark masses, for trilinear
couplings of electroweak gauge strength~\cite{Adloff:2003jm}. Similar
bounds, under the same assumptions, have been found at LEP from their
search for anomalous 4-fermion vertices~\cite{Abbiendi:1998ea}. For first
generation leptoquark trilinear  couplings which are much smaller than
gauge strength (as is the case for the Yukawas in the SM), the mass bounds
are greatly weakened.

At the Tevatron, the leptoquarks could be produced in pairs, with
identification made through decay topologies. In this way, the bounds
are not dependent on the trilinear couplings, except for decay
branching fractions. In the case of first~\cite{Abbott:1997mq} and
second~\cite{Abazov:2006vc} generation leptoquarks, the final state
topology consists of 2 hadronic jets + 2 charged leptons, and the
resulting lower limits on the leptoquark mass are around 250 GeV.  In
the case of the third generation, a lower limit of 219 GeV has been
recently reported by the D\O~Collaboration~\cite{D0}, by tagging on 2
$b$ jets + missing energy.  For this value of the leptoquark mass, the
decay into $t \tau$ is largely suppressed compared to the $b \nu_\tau$
channel, so that the mass bound is nearly independent of even the
branching fraction. As the explored mass region becomes larger, the $t
\tau$ channel becomes more available and thus the mass limit obtained
is pushed a bit lower (to $\ge$ 213 GeV).

In this work, we explore the possibility of probing
second and third generation leptoquark parameter space (coupling and
mass) with the IceCube neutrino detection
facility~\cite{Ahrens:2002dv}. This experiment, located below the
surface of the Antarctic ice sheet at the geographic South pole, is
required to be sensitive to the best estimates of potential cosmic ray
neutrino fluxes.  When completed, the telescope will consist of 80
kilometer-length strings, each instrumented with 60 10-inch
photomultipliers spaced by 17~m.  The deepest module is 2.4~km below
the surface. The strings are arranged at the apexes of equilateral
triangles 125\,m on a side. The instrumented (not effective!)
detector volume is a cubic kilometer. IceTop, a surface array of
{C}erenkov detectors deployed over 1~km$^{2}$ above IceCube, augments
the deep-ice component by providing a tool for calibration, background
rejection and air-shower physics.  The expected energy resolution is
$\pm 0.1$ on a log$_{10}$ scale.  Construction of the detector started
in the Austral summer of 2004/2005 and will continue for 6 years,
possibly less. At present, data collection by the first
9 strings has begun.

The event signatures are grouped as tracks, showers, or a combination
of the two. Tracks include muons resulting from both cosmic ray
showers and from charged current (CC) interaction of muon neutrinos.
Tracks can also be produced by $\tau$ leptons arising in ultra-high
energy $\nu_\tau$ CC interactions. Showers are generated by neutrino
collisions $(\nu_e$ or low energy $\nu_\tau$ CC interactions, and all
neutral current interactions) inside or near the detector, and by
muon bremsstrahlung radiation near the detector.

The experimental situation is greatly simplified for neutrino energy
$E_\nu \agt 10^6~{\rm GeV}.$ A cut at this energy is sufficient to reduce
the great majority of background from muon bremsstrahlung and tracks
arising from muons produced in cosmic ray showers. Moreover, the flux of
atmospheric neutrinos is low above this energy~\cite{Lipari:1993hd}, so
this cut generates a very pure sample of extraterrestrial neutrinos. Of
particular interest here, for $E_\nu
> 10^{6}~{\rm GeV}$, there is sufficient energy resolution
  ($\pm 0.2$ on a log$_{10}$ scale) to separately assign the
energy fractions in
the muon track and the hadronic shower, allowing the determination of
the inelasticity distribution. Similarly, in the energy decade
$10^{6.5} < E_\nu/{\rm GeV} < 10^{7.5}$ one can expect good resolution
(less than 5\%) in ``double bang'' events generated
by incoming $\nu_\tau$'s. Again, this will allow a reasonable
measurement of the inelasticity distribution.

In this study, we emphasize the inelasticity $(y)$
distribution of events as an important tool for detection of new
physics. In particular, we will find that the $y$ distribution of
events generated through resonant leptoquark production differs
substantially from the SM prediction. If the event rate for the
new physics turns out to be comparable to SM expectations, then
the $y$ profile of the measured data can be used to probe the
coupling-mass leptoquark parameter space. The outline of the paper
is as follows: In Sec.~\ref{2} we derive the relevant
$y$-distribution of events generated through production and decay
of a scalar leptoquark under the assumption of diagonal generation
coupling. Armed with this distribution, in Sec.~\ref{3} we present
a statistical method for assessing the significance of discovery
criteria. Our conclusions are collected in Sec.~\ref{4}.

\section{Leptoquark Phenomenology}
\label{2}

A general Lagrangian for $SU(3)_C\otimes SU(2)_L\otimes U(1)_Y$-invariant
flavor-diagonal leptoquark couplings has been presented
in~\cite{Buchmuller:1986zs}. To illustrate our proposal, we consider the
simple case of SU(2)-singlet scalar leptoquarks $S_i$ which interact
with quarks and leptons through the Lagrangian
\begin{equation}
\mathfrak{L}_{LQ} = \sum_i (g_L \,\,{\overline Q}^{\;c}_{iL} \,\,i \tau_2 \,\,
L_{iL} + g_R \,\, {\overline u}^{\;c}_{iR}\,\, l_{iR})\, S_i \ .
\end{equation}
Here $Q_i = (u_i\ d_i)^T$ and $L_i = (\nu_i\ l_i)^T$ stand for quark and
lepton SU(2) left-handed doublets, $u_{iR}$ and $l_{iR}$ are right-handed
singlets, and $g_L$ and $g_R$ are the corresponding coupling constants.
Subindices $i$, running from 1 to 3, label the quark or fermion family.
For simplicity we will assume that the interaction conserves leptoquark
family quantum numbers separately (i.e., there is no mixing between
different families). Thus in the following subindices $i$ will be dropped,
and up- and down-like quarks will be denoted generically with $U$ and $D$
respectively.

\begin{figure}
 \postscript{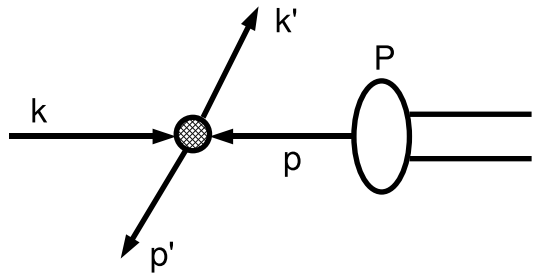}{0.98}
 \caption{Schematic diagram of a neutrino-parton collision, in which a
neutrino with momentum $k$ hits a quark with momentum $p = x P$ giving
rise to a secondary charged lepton and quark with momenta $k'$ and
$p',$ respectively. Here $x$ is the fractional energy of the struck
parton in the nucleon having momentum $P.$}
\label{fig:1}
\end{figure}

We will be considering the inclusive $\nu\, P$ scattering schematically
shown in Fig~\ref{fig:1}. The cross-hatched circle includes both resonant
leptoquark production and decay, as well as $u$-channel exchange of a
leptoquark leading to the same final state. We will not be considering
couplings $g_{L (R)} \agt 2,$ since such a coupling could lead to Landau
singularities at low energies.  Hence, we will assume that the resonant
cross section largely dominates the process, and the narrow width
$\delta$-function approximation will be valid.

Let us assume that an incoming neutrino collides with a proton target with
center-of-mass energy $s$. If the neutrino hits a down-like parton $D$,
the inclusive cross section for the process shown in Fig.~\ref{fig:1} in
the parton model is given by
\begin{equation}
\frac{d\sigma_{LQ} (\nu P \to l^- X)}{d^3k'/E'}  = \int dx \,D(x)\,
\frac{d\hat\sigma_{LQ} (\nu D \to l^- U)}{d^3k'/E'} \ ,
\label{sigma}
\end{equation}
where $D(x)$ is the corresponding parton distribution function (pdf).
The neutrino-parton cross section reads
\begin{widetext}
\begin{equation}
\frac{d\hat \sigma_{LQ}(\nu D \to l^- U)}{d^3k'/E'}  =
\frac{1}{(2 \pi)^2}\,
\frac{1}{2{\cal F}}
 \int d^4p'\,\, \delta^4(k+p-k'-p')\,\,\,
\delta^+(p'^2 - m_u^2)\,\, \frac{1}{2}
\,\sum_{\rm spin}\, \left(|\mathfrak{M}_L|^2 + |\mathfrak{M}_R|^2
\right)\ ,
\label{hatcs}
\end{equation}
\end{widetext}
where ${\cal F} = 2xs = 2\hat s$ is the invariant flux, $m_U$ is the mass
of the outgoing up-like quark, and $E'$ is the lab energy of the outgoing
charged lepton. In the resonant approximation, the amplitude for the
production of a left-handed charged lepton is given by
\begin{equation}
\mathfrak{M}_L = g_L^2 \ \frac{\overline l_L(k')\, U^c_L (p')\,
\overline D^{\;c}_L(p)\, \nu_L(k)}{\hat s- M^2- i
\Gamma M} \, ,
\end{equation}
where $M$ and $\Gamma$ are the mass and width of the leptoquark. A similar
expression, replacing $g_L^2\to g_L\, g_R$, holds for the decay through
the right-handed channel. It is easy to see that there is no interference
between $\mathfrak{M}_L$ and $\mathfrak{M}_R$.

Now in the narrow resonance approximation one has
\begin{equation}
 \frac{1}
{(\hat s - M^2)^2 + (\Gamma M)^2}  \ \to \
\frac{\pi}{\Gamma M}\,\, \delta(\hat s - M^2)\ \ .
\end{equation}
{}Then, after summing over spins of outgoing fermions one arrives to
\begin{eqnarray}
\!\!\!\frac{1}{2} \sum_{\rm spin}\,
\left(|\mathfrak{M}_L|^2 + |\mathfrak{M}_R|^2\right) & = & \frac{\pi}{2}
\, g_L^2\, (g_L^2+g_R^2) \nonumber \\
& \times & \, \frac{\hat s\;(\hat s - m_U^2)}{M\,\Gamma}\;
\delta(\hat s - M^2)\; ,
\label{sqamp}
\end{eqnarray}
where we have neglected both the mass of the down-like quark and the mass
of the outgoing charged lepton. We have kept instead the $m_U$ dependence,
since it is relevant in the case of the third family, where the outgoing
quark would be a top. As a further assumption, we will consider that the
leptoquark width is dominated by the $Ul$ and $D\nu$ quark-lepton
channels. This leads to
\begin{equation}
\Gamma \ = \ \frac{M}{16 \pi} \bigg\{
g_L^2\; [\,(1-\lambda_U)^2+1\,] + g_R^2\; (1-\lambda_U)^2\bigg\}\ ,
\label{gamma}
\end{equation}
where $\lambda_U = m_U^2/M^2$. Substituting Eqs.~(\ref{hatcs}),
(\ref{sqamp}) and (\ref{gamma}) into Eq.~(\ref{sigma}) yields
\begin{eqnarray}
\frac{d\sigma_{LQ}^{(\alpha)}}{d^3k'/E'} & = & \frac{g_L^2\,(g_L^2 +
g_R^2)(1-\lambda_U)} {(g_L^2 + g_R^2)(1-\lambda_U)^2+g_L^2}\,\, \int
dx\,\,
\frac{D(x)}{2xs}\nonumber \\
 & \times & M^2\ \delta(Q^2 - 2 m_p \,y\,E_\nu\, x + m_U^2)\,
\delta(xs - M^2) \, \nonumber \\
& = & \frac{1}{2\,s}\,
\frac{g_L^2\,(g_L^2 + g_R^2)(1 - \lambda_U)}
{(g_L^2 + g_R^2)(1-\lambda_U)^2+g_L^2}
\nonumber \\ & \times & D(M^2/s)\;
\delta(Q^2 - y\, M^2+m_U^2) \ ,
\label{sm}
\end{eqnarray}
where $Q^2 = - (k-k')^2$, and the inelasticity $y$ is defined as $y =
(E_\nu - E')/E_\nu$, $E_\nu$ being the lab energy of the incoming
neutrino. Indices $\alpha =1,2,3$ correspond to the up-like quarks
$U=u,c,t$, respectively. After adequate changes of variables and
integrations~\cite{Halzen:1984mc} we find
\begin{eqnarray}
\!\!\!\frac{d \sigma_{LQ}^{(\alpha)}}{dy} &= & \frac{\pi}{2}\;
\frac{g_L^2\,(g_L^2 + g_R^2)(1-\lambda_U)}
{(g_L^2 + g_R^2)(1-\lambda_U)^2+g_L^2}\;
\frac{D(M^2/s)}{s} \ .
\label{cuqui}
\end{eqnarray}
The inelasticity $y$ lies in the range $\lambda_U\leq y \lesssim 1$. Note
that the $y$ distribution of the resonant process is approximately flat
(at the energies of interest, the $Q^2$ dependence of the pdf can be
neglected), in contrast to that characterizing SM charged current (CC)
processes in which
\begin{eqnarray}
\frac{d\sigma_{\rm SM}^{\rm CC}}{dy} & = &
\frac{2\,G_{\rm F}^2 \, m_p E_\nu}{\pi} \,\,
\left(\frac{M_W^2}{Q^2 + M_W^2}\right)^2 \nonumber \\
 & \times & \int dx [xq(x, Q^2) + x \overline q(x, Q^2) (1-y)^2] \,,
\label{smJ}
\end{eqnarray}
where $G_{\rm F} = 1.16632 \times 10^{-5}~{\rm GeV}^{-2}$ is the Fermi
constant, $M_W$ is the mass of the $W$ gauge boson, and $q(x)$ and
$\overline q (x)$ stand for combinations of quark and anti-quark
proton pdf's, respectively~\cite{Gandhi:1998ri}. The $y$ dependence of
the SM cross section is shown in Fig.~6 of Ref.~\cite{Gandhi:1995tf}.
In the next section we exploit the differing $y$ dependences of the
leptoquark and SM interactions to constrain the parameter space of the
new physics.

\section{Sensitivity Reach at IceCube}
\label{3}

To evaluate the prospects for probing leptoquark production at IceCube,
one has to estimate the ``beam luminosity'', i.e.\ the magnitude of the
(yet to be detected) neutrino flux. We know that cosmic accelerators
produce particles with energies in excess of $10^{11}~{\rm GeV}$ (we do
not know where or how~\cite{Anchordoqui:2002hs}), and a neutrino beam is
expected to come in association with these cosmic
rays~\cite{Gaisser:1994yf}. However, given our ignorance of the opacity of
the sources, it is difficult to calculate the magnitude of the neutrino
flux. The usual benchmark here is the so-called Waxman-Bahcall (WB) flux
\begin{equation}
  E_\nu^2\; \phi^\nu_{\rm WB}(E_\nu) \ \simeq \
  6 \times 10^{-8}~{\rm GeV}\, {\rm cm}^{-2} \,
{\rm s}^{-1} \,{\rm sr}^{-1}
\end{equation}
(all flavours), which is derived assuming that neutrinos come from
transparent cosmic ray sources~\cite{Waxman:1998yy}, and that there is
adequate transfer of energy to pions following $pp$ collisions. Here we will
rely on this expression to estimate the event rates needed to quantify the
IceCube sensitivity to leptoquark production.  However, one should keep in
mind that if there are in fact ``hidden'' sources which are opaque to
ultra-high energy cosmic rays, then the expected neutrino flux will be
higher~\cite{Stecker:1991vm}. Moreover, if the extragalactic cosmic rays
begin to dominate over the Galactic component at energies as low as $\sim
10^9$ GeV, as suggested recently~\cite{Berezinsky:2002nc}, then the
required power of the extragalactic sources will increase by a factor of
$\sim 2$, implying a concommitantly larger neutrino
flux~\cite{Ahlers:2005sn}.

IceCube is sensitive to both downward and upward coming cosmic neutrinos.
However, to remain conservative with our statistical sample, here we
select only downward going events. To a good approximation, the expected
number of such events at IceCube is given by
\begin{equation}
{\cal N} = 2 \pi\, n_{\rm T}\, T\, \int dE_\nu\,\,
\sigma_{\rm tot} (E_\nu)\,\, \phi^\nu_{\rm WB} (E_\nu)\ ,
\label{eventrate}
\end{equation}
where $n_{\rm T}$ is the number of target nucleons in the effective
volume, $T$ is the running time, and $\sigma_{\rm tot} (E_\nu)$ is the
total neutrino-nucleon cross section. In our analysis we are interested
only in CC contained events, for which an accurate measurement of the
inelasticty can be obtained. The IceCube's effective volume for
(background rejected) contained events is roughly $1~{\rm
  km}^{3}$~\cite{Anchordoqui:2005is}, which corresponds to $n_{\rm T}
\simeq 6 \times 10^{38}.$

\begin{figure}
 \postscript{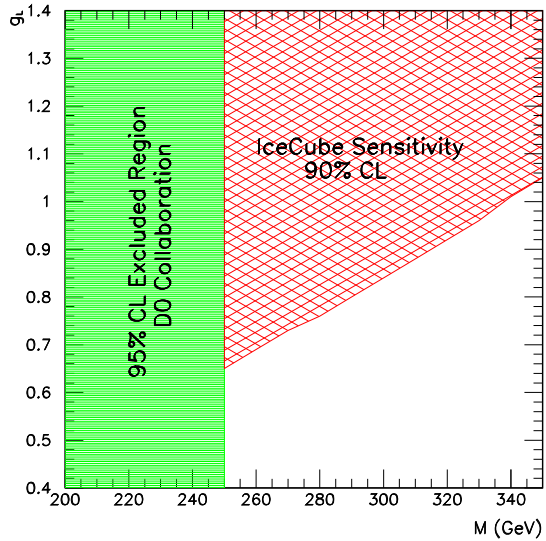}{0.8}
   \caption{Sensitivity reach of IceCube (90\% CL) to probe second
     generation SU(2)-singlet scalar leptoquark parameter space
     (coupling and mass).  For comparison, the existing limit (95\%
     CL) reported by D\O Collaboration~\cite{Abazov:2006vc} is also
     shown.}
\label{fig:2}
\end{figure}

Hereafter we focus on neutrino energies in the range $10^7 < E_\nu/{\rm
GeV} < 10^{7.5}$, where the background from atmospheric neutrinos is
negligible, but the extraterrestrial flux is expected to be significant.
Thus, we will consider a medium energy $\langle E_\nu \rangle =
10^{7.25}$~GeV. At production, the WB flux has flavor ratios
$\nu_\mu:\nu_e:\nu_\tau = 2: 1: 0$, but this quickly transforms to $1:1:1$
through neutrino oscillations~\cite{Learned:1994wg}. One has then
\begin{equation}
\phi_{\rm WB}^{\nu_\alpha} (\langle E_\nu \rangle) \ \simeq \
6 \times 10^{-23}~{\rm GeV}^{-1}~{\rm cm}^{-2}~{\rm s}^{-1}~{\rm sr}^{-1}
\label{flux}
\end{equation}
for each flavor $\alpha$. Now it is possible to increase the ratio
signal/background events by performing a cut in the inelasticity $y$.
Given the dependence on $y$ of the $\sigma_{\rm SM}^{\rm CC}$ cross
section, the flat behavior of the $\sigma_{LQ}$ cross section, and
the available phase space for quark production, it is convenient to
consider events with relatively large values of $y$. We choose here
events in the range $y \geq 0.5$. With this cut, the integration of
Eq.~(\ref{smJ}) leads to
\begin{equation}
\left. \sigma_{\rm SM}^{\rm CC} (\langle E_\nu \rangle)
\right|_{y\,\geq\, 0.5} \
\simeq \ 8 \times 10^{-34} \ {\rm cm}^2 \ .
\end{equation}
One has to take into account that this result carries a systematic
error of about 20\%~\cite{Gandhi:1998ri,Frichter:1994mx} due to
uncertainties in the extrapolation of the pdf's~\cite{Pumplin:2002vw}.
For illustrative comparison the second generation leptoquark cross
section in the high $y$ region is calculated from Eq.~(\ref{cuqui}) to
be
\begin{equation}
\left.
\sigma_{LQ} (\langle E_\nu \rangle)\right|_{y\, \geq\, 0.5} \simeq \ 2 \times 10^{-33} \ {\rm cm}^2 \ ,
\end{equation}
for fiducial values $g_L = g_R = 1$ and leptoquark mass $M = 300~{\rm GeV}$.
From
Eqs.~(\ref{eventrate}) and (\ref{flux}) we can now easily estimate the
number of expected SM background events during the lifetime of the
experiment. Taking $T = 15$~yr, for each neutrino flavor $\alpha$ one has
\begin{eqnarray}
{\cal N_{\rm B}}^{(\alpha)} & \simeq & 2 \pi \,n_{\rm T} \,T \,\left.
\sigma_{\rm SM}^{\rm CC} (\langle E_\nu \rangle)\right|_{y\,\geq\, 0.5}
\; \phi^{\nu_\alpha}_{\rm WB} (\langle E_\nu \rangle)\,\,
\Delta E_\nu \nonumber \\
   & = & 2 \,\,,
\end{eqnarray}
where $\Delta E_\nu = 10^{7.5}~{\rm GeV} - 10^7~{\rm GeV} \simeq 2.2
\times 10^7~{\rm GeV}$. (Note that the background is cosmic neutrinos!
Todays' signal, tomorrow's background.) In the same way, the number of
signal events ${\cal N}_{\rm S}$ will be approximately given by
\begin{equation}
{\cal N_{\rm S}}^{(\alpha)} \simeq  2 \pi\, n_{\rm T}\, T\, \left.
\sigma_{LQ} (\langle E_\nu \rangle)\right|_{y\, \geq\, 0.5} \;
\phi^{\nu_\alpha}_{\rm WB} (\langle E_\nu \rangle) \,\, \Delta E_\nu \ ,
\end{equation}
which for the above proposed leptoquark interaction is just a function of
the couplings $g_{R,L}$ and the leptoquark mass $M$.

\begin{figure}[hbt]
 \postscript{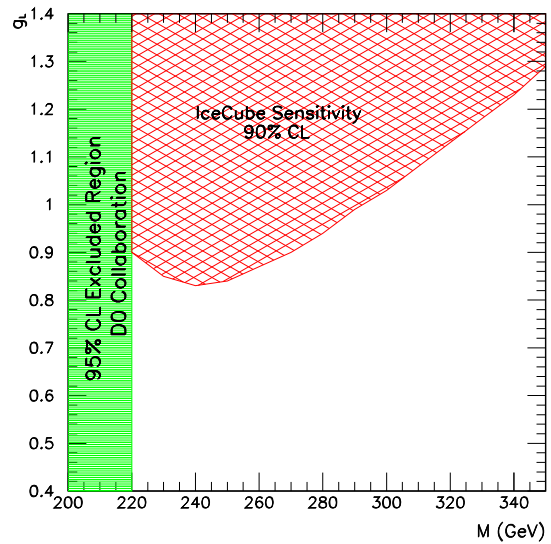}{0.8}
 \caption{Sensitivity reach of IceCube (90\% CL) to probe third generation
SU(2)-singlet scalar leptoquark parameter space (coupling and mass). For
comparison, the existing limit (95\% CL) reported by D\O
Collaboration~\cite{D0} is also shown.} \label{fig:3}
\end{figure}

It should be noted that the neutrino induced events do not constitute
the sole background. As mentioned in the introduction there are muons
(produce in the atmosphere) which traverse the detector and may
deposit energy through  bremsstrahlung radiation. In our energy bin, one may
expect 10 muon traversals in 15 yr. However, our inelasticity cut will
completely eliminate this source of background, because of the
negligible probability for muons to radiate 50\% of their energy.

To determine the bounds for leptoquark production, let us assume that 2
$\nu_{\alpha}$-events are in fact observed with $y \geq 0.5$. Then, at
90\% CL, there will be an upper bound on signal events given by ${\cal
N}_{\rm S}^{(\alpha)}\leq 3.91$~\cite{Feldman:1997qc}. For simplicity, we
consider the left-right symmetric case in which $g_L = g_R$. Then, after
numerical evaluation of the leptoquark cross sections, the upper bounds on
${\cal N}_{\rm S}^{(\alpha)}$ can be translated into contours of constant
likelihood in the $M$-$g_L$ plane. Our results are displayed in
Figs.~\ref{fig:2} and \ref{fig:3}, where we show the sensitivity reach of
IceCube together with the existing limits from D\O
Collaboration~\cite{Abazov:2006vc,D0}.

In the case of the third family, it can be seen that the sensitivity is
maximal for leptoquarks of $M \simeq 245$~GeV. For lower leptoquark masses
(in the narrow resonance limit) the allowed inelasticity range ---and thus
the leptoquark cross section--- becomes reduced due to phase space suppression,
owing to the large mass of the top quark.

In order to estimate the significance of the assumption $g_L = g_R$,
we have also considered the case of purely left-handed leptoquark
currents, i.e.\ $g_R=0$. By looking at Eq.~(\ref{cuqui}), it can be
seen that this implies an average reduction in the leptoquark cross
section by a factor of about 0.75 and 0.65 for the second and third
families, respectively.

\section{Conclusions}
\label{4}

In this paper we have introduced the measurement of inelasticity as
a powerful tool for probing new physics in cosmic neutrino interactions.
As an illustrative example, we have discussed the possibility of detecting
leptoquark production at the IceCube neutrino
telescope~\cite{Robinett:1987ym}. We estimated the
expected event rate at IceCube to be comparable to the one
predicted for cosmic ray facilities that make use of the atmosphere as the
detector calorimeter~\cite{EspiritoSanto:2005mx}. However, the ability of
IceCube to accurately measure the inelasticity distribution of events
provides a unique method for SM background rejection, allowing powerful
discrimination of resonant processes: we have shown that production of
leptoquarks with masses $\agt 250$~GeV and diagonal generation couplings
of ${\cal O}(1)$ can be directly tested at the Antarctic ice-cap.

In closing, some comments are in order. First, we have not taken
account of any systematic considerations concerning the detector --
these are beyond the scope of the present work. Second, for reasons of
simplicity, we have not included upcoming events close to the horizon.
Although these Earth-skimming neutrinos have the potential of nearly
doubling our signal event background (and thus nearly halving the
required observation time scale), their proper consideration will
require a full Monte Carlo simulation.

\acknowledgments{CAGC is supported by CONICET and ANPCyT, Aregentina.
  HG is supported by the US National Science Foundation (NSF) Grant No.
  PHY-0244507.  DGD is supported by CONICET and ANPCyT, Argentina, under
  grants PIP 6009, PICT04 03-25374 and PICT02-03-10718.  FH is supported
  in part by the US NSF under Grant No. OPP- 0236449, in part by the US
  Department of Energy (DoE) Grant No. DE-FG02-95ER40896, and in part by
  the University of Wisconsin Research Committee with funds granted by the
  Wisconsin Alumni Research Foundation.}

\end{document}